\documentclass[11pt]{article}
\usepackage{amssymb,amsmath,mathrsfs,enumerate}
\usepackage{graphicx,rotate,multicol}
\usepackage[margin=10pt,labelfont=bf]{caption}
\usepackage{cite}
\usepackage[utf8]{inputenc}
\usepackage[colorlinks=true,
            linkcolor=red,
            urlcolor=blue,
        citecolor=blue]{hyperref}

\usepackage{color}
\usepackage{lineno}
\long\def\rpl#1!!!#2!!!{\color[rgb]{.7,0,0}{#1} \color{blue}{#2} \color{black}}

\let\tilde=\widetilde

\let\bar=\overline

\def \order(#1){{\mathcal O} \left(#1 \right)}
\def\rep#1{\ensuremath{\mathbf #1}}

\evensidemargin=\oddsidemargin
\def\Eqn#1{Eq.\ (\ref{#1})}
\def\Eqs#1#2{Eqs.\ (\ref{#1}) and (\ref{#2})}
\allowdisplaybreaks

\title	{\LARGE\bf 
Quark mixing in an $S_3$ symmetric model \\ with two Higgs doublets}

\author {\sf Dipankar
  Das,$^{a,}$\footnote{ddphy@caluniv.ac.in} \quad  Ujjal 
  Kumar Dey,$^{b,}$\footnote{ujjal@cts.iitkgp.ernet.in} \quad Palash
B. Pal$^{c,}$\footnote{pbpal@theory.saha.ernet.in } \\[10pt]
\small\em $^a$Department of Physics, University of Calcutta, 92
Acharya Prafulla Chandra Road, Kolkata 700009, India\\ 
\small\em $^b$Centre for Theoretical Studies, Indian Institute of
Technology Kharagpur, Kharagpur 721302, India \\ 
\small\em $^c$Theory Division,
Saha Institute of Nuclear Physics, 1/AF Bidhan Nagar, Kolkata 700064,
India \\ }

\date{}

\begin{document}


\maketitle	

\begin{abstract}
We construct a model where the smallness of the masses of first quark
generations implies the near block diagonal nature of the CKM matrix
and vice-versa.  For this set-up, we rely on a 2HDM structure with an
$S_3$ symmetry.  We show that an SM-like Higgs emerges naturally from
such a construction.  Moreover, the ratio of two VEVs, $\tan\beta$ can
be precisely determined from the requirement of the near masslessness
of the up- and down-quarks.  The FCNC structure that arises from our
model is also very predictive.
\end{abstract}

\bigskip

The Standard Model (SM) does not provide any connection between quark
masses and mixings: they are independent parameters to be fixed by the
experimental observations.  One attractive way to obtain insight into
these parameters is to impose some additional symmetry under which the
generations of quarks transform in a non-trivial way.  There have been
many attempts where, by imposing a discrete symmetry on different
generations of SM fermions, some relations between the masses and
mixings have been obtained (see \cite{Altarelli:2010gt,
  Ishimori:2010au} for review).

In this article, we present an attempt to relate two features of quark
masses and mixings.  The first of these two is the fact that the first
generation quarks are very light compared to the other ones whereas
the second concerns the near block-diagonal structure of the quark
mixing matrix, or the Cabibbo-Kobayashi-Maskawa~(CKM) matrix.  This
second feature comes out very clearly in the Wolfenstein
parametrization \cite{Wolfenstein:1983yz} of the CKM matrix, where
each element is written in a power series of a small parameter
$\lambda$.  If we keep only terms up to the linear order in $\lambda$,
the CKM matrix is indeed block-diagonal.  We propose a connection
between these two features by invoking an $S_3$ symmetry.

Many works on flavour model building using $S_3$ symmetry have been
done in the past~\cite{Pakvasa:1977in, Koide:1999mx, Kubo:2003iw,
  Harrison:2003aw, Chen:2004rr, Koide:2006vs, Mondragon:2007af,
  Chen:2007zj, Jora:2009gz, Kaneko:2010rx, Teshima:2011wg, Dev:2011qy,
  Dias:2012bh, Meloni:2012ci, Dev:2012ns, Zhou:2011nu, Ma:2013zca,
  Benaoum:2013ji, Ma:2014qra, Das:2015sca, Cogollo:2016dsd,
  Pramanick:2016mdp, Varzielas:2016zuo}.  In these constructions one
usually employs, for the scalar sector, a three Higgs doublet
structure which goes well with the aesthetic idea of having three
replicas of Higgs doublets in conformity with three generations of
fermions~\cite{Deshpande:1991zh, Kubo:2004ps, Teshima:2005bk,
  Koide:2005ep, Beltran:2009zz, Ferreira:2010hy, Bhattacharyya:2010hp,
  Bhattacharyya:2012ze, Teshima:2012cg, Machado:2012ed,
  Barradas-Guevara:2014yoa, Das:2014fea, Barradas-Guevara:2015rea,
  Merchand:2016ldu, Emmanuel-Costa:2016vej}.  Even more complicated
scalar structures are not uncommon~\cite{Xing:2010iu, Canales:2012dr,
  Canales:2013cga, Hernandez:2014vta, Hernandez:2015dga,
  Hernandez:2015zeh, Gomez-Izquierdo:2017rxi}.
However, in this paper, we rely on a two Higgs-doublet model~(2HDM)
scalar structure~\cite{Branco:2011iw} which is much more economical in
terms of independent parameters.  Although the idea of a 2HDM with
$S_3$ symmetry has been conceived lately \cite{Cogollo:2016dsd}, some
distinct implications have not been emphasised earlier.  For example,
we will show that an $S_3$ symmetric 2HDM potential naturally delivers
an SM-like Higgs boson which can be identified with the scalar
resonance observed at the LHC with signal strengths in close agreement
with the SM predictions~\cite{Khachatryan:2016vau}.  We will also
demonstrate how, in our scenario, the requirement of near-masslessness
for the first generation of quarks dictates a particular value of
$\tan \beta$, which will simultaneously render the CKM matrix
block-diagonal.  For intuitive understanding of the model Lagrangian
and the conclusions that follow from it, a brief overview of the $S_3$
symmetry is in order.

The discrete symmetry group $S_3$ has three irreducible
representations: \rep1, \rep{1'} and \rep2.  We pick a basis such that
the generators in the ${\bf 2}$ representation are given by
\begin{eqnarray}
a = \begin{bmatrix} -\frac12 & \frac{\surd3}2 \\
- \frac{\surd3}2 & -\frac12
\end{bmatrix} \,, 
\qquad
b = \begin{bmatrix}\frac12 &
  \frac{\surd3}2  \\ \frac{\surd3}2 & - \frac12 
  \end{bmatrix}\,.
\label{gen}
\end{eqnarray}
Note that $a$ is of order 3, whereas $b$ is of order 2.  The rest of
the elements can be obtained by taking products of powers of these two
elements.  In this basis the quark fields transform under $S_3$ in
the following way:
\begin{subequations}
\label{quarkS3}
\begin{eqnarray}
{\bf 2}   & : & \; \begin{bmatrix}Q_1 \\ Q_2 \end{bmatrix},
\; \begin{bmatrix}u_{1R} \\ u_{2R} \end{bmatrix},
\; \begin{bmatrix}d_{1R} \\ d_{2R} \end{bmatrix}\,, 
\label{qLS3} \\*
 {\bf 1} & : & \; Q_3, \; u_{3R}, \; d_{3R} \,,
\label{qRS3}
\end{eqnarray}
\end{subequations}
where the $Q_A$'s ($A=1,2,3$) are the usual left-handed $\rm SU(2)$
quark doublets, whereas the $u_{AR}$'s and $d_{AR}$'s are the
right-handed up-type and down-type quark fields respectively, which
are singlets of the $\rm SU(2)$ part of the gauge symmetry.  Note that
the square brackets, in \Eqs{gen}{quarkS3} as well as in the
subsequent text, denote the doublet representation of $S_3$, and has
nothing to do with the representation of the enclosed fields under
$\rm SU(2)$.  Similarly, in the Higgs sector, there are two $\rm SU(2)$
doublets $\phi_i$ ($i=1,2)$, and their transformation under the $S_3$
symmetry is as follows:
\begin{eqnarray}
{\bf 2} & : & \begin{bmatrix} \phi_1 \\ \phi_2 \end{bmatrix} \equiv \Phi
\,.  
\end{eqnarray}
We write the potential of the theory as follows:
\begin{subequations}
\label{potential}
\begin{eqnarray}
V(\Phi)&=& V_2(\Phi) +V_4(\Phi) \,, 
\end{eqnarray}
where $V_2(\Phi)$ contains only quadratic terms in the fields $\phi_1$
and $\phi_2$ whereas $V_4(\Phi)$ contains quartic terms.  The most
general $S_3$-invariant form for $V_4(\Phi)$ is:
\begin{eqnarray}
V_4(\Phi) &=& \lambda_1 (\phi_1^\dagger\phi_1+\phi_2^\dagger\phi_2)^2
+\lambda_2 (\phi_1^\dagger\phi_2 -\phi_2^\dagger\phi_1)^2 \nonumber\\*
&& + \lambda_3
\left\{(\phi_1^\dagger\phi_2+\phi_2^\dagger\phi_1)^2
+(\phi_1^\dagger\phi_1-\phi_2^\dagger\phi_2)^2\right\}\,.  
\label{V4}
\end{eqnarray}
For the quadratic terms, the most general form is given by
\begin{eqnarray}
V_2(\Phi) &=& \mu_1^2(\phi_1^\dagger\phi_1)
       + \mu_2^2(\phi_2^\dagger\phi_2) 
- \Big( \mu_{12}^2 \phi_1^\dagger\phi_2 + \mbox{h.c.} \Big)  \,,
\label{V2}
\end{eqnarray}
\end{subequations}
which is not $S_3$-symmetric unless the co-efficients satisfy some
special conditions.  If these conditions are not met, $V_2(\Phi)$
contains terms which softly break the $S_3$ symmetry, and we allow for
such terms.  We will consider various scenarios with the quadratic
terms in a short while.

The parameters in the quartic part of the potential must be real
because of hermiticity of the Lagrangian.  In the quadratic part
$V_2(\Phi)$, the parameters $\mu_1^2$ and $\mu_2^2$ are also real.
The parameter $\mu_{12}^2$ can be complex, but its phase can be
absorbed by redefining either $\phi_1$ or $\phi_2$.  Thus, all
parameters in $V(\Phi)$ can be taken to be real without any loss of
generality.  It has been argued \cite{Ferreira:2010hy} that in this
case the vacuum expectation values~(VEVs) can also be taken to be
real.  Denoting the VEV of $\phi_{i}$ by $v_{i}$ we write the doublets
after symmetry breaking in the form
\begin{eqnarray}
\phi_i = {\phi_i^+ \choose \frac1{\surd2} (v_i + h_i + i \zeta_i) }
\,, 
\label{phii}
\end{eqnarray}
and use the standard notation
\begin{eqnarray}
v_1 = v \cos\beta\,, \qquad v_2 = v \sin\beta \,,
\end{eqnarray}
where the $W$ and $Z$-boson masses are proportional to $v\approx
246$~GeV.  Assuming both $v_1$ and $v_2$ to be non-zero, the
minimisation condition of the potential $V(\Phi)$ can be written as 
\begin{subequations}
\label{min}
\begin{align}
\mu_1^2 &= \mu_{12}^2 \tan\beta - (\lambda_1 + \lambda_3) v^2 \,, \\*
\mu_2^2 &= \mu_{12}^2 \cot\beta - (\lambda_1 + \lambda_3) v^2 \,.
\end{align}
\end{subequations}

Let us discuss the physical scalar spectrum of the model.  We begin
with the charged boson sector.  One combination of $\phi_1^\pm$ and
$\phi_2^\pm$, to be denoted by $w^\pm$, will constitute an unphysical
field that does not appear in the physical spectrum.  The orthogonal
combination, $H^\pm$, will be a physical charged scalar.  The two
combinations will be given by
\begin{eqnarray}
\begin{pmatrix}
w^\pm \\ H^\pm
\end{pmatrix}
= \begin{pmatrix}
\cos\beta & \sin\beta \\
-\sin\beta & \cos\beta 
\end{pmatrix}
\begin{pmatrix}
\phi_1^\pm \\ \phi_2^\pm
\end{pmatrix} \,.
\end{eqnarray}
The mass of the physical charged scalar can be easily calculated:
\begin{eqnarray}
M_{H^\pm}^2 = {2 \mu_{12}^2 \over \sin2\beta} - 2\lambda_3 v^2 \,.
\end{eqnarray}
In the pseudoscalar sector, there is one combination, $z$, which
becomes unphysical after symmetry breaking, and there is one physical
pseudoscalar field $A$.  They are given by
\begin{eqnarray}
\begin{pmatrix}
z \\ A
\end{pmatrix}
= \begin{pmatrix}
\cos\beta & \sin\beta \\
-\sin\beta & \cos\beta 
\end{pmatrix}
\begin{pmatrix}
\zeta_1 \\ \zeta_2
\end{pmatrix} \,
\end{eqnarray}
with
\begin{eqnarray}
M_A^2 = {2\mu_{12}^2 \over \sin 2\beta} - 
                    2(\lambda_2 + \lambda_3) v^2 \,.
\end{eqnarray}
The mass matrix for the scalar part can be written as,
\begin{align}
V_{\rm mass}^{\rm S} = 
\begin{pmatrix}
h_1 & h_2
\end{pmatrix}
\frac12 {\mathbb M}_S^2
\begin{pmatrix}
h_1 \\ h_2
\end{pmatrix}
\end{align}
with
\begin{align}
{\mathbb M}_S^2 = \begin{pmatrix}
\mu_{12}^2\frac{v_2}{v_1} + 2v_1^2
(\lambda_1 + \lambda_3) & 
-\mu_{12}^2 + 2v_1v_2(\lambda_1 + \lambda_3) \\
-\mu_{12}^2 + 2v_1v_2(\lambda_1 + \lambda_3) &
\mu_{12}^2\frac{v_1}{v_2} + 2v_2^2
(\lambda_1 + \lambda_3)
\end{pmatrix}.
\end{align}
The diagonalisation of ${\mathbb M}_S^2$ will lead to two neutral physical
scalars $H$ and $h$,
\begin{align}
\label{eq:cpevn}
\begin{pmatrix}
h \\ H
\end{pmatrix} = 
\begin{pmatrix}
\cos\beta & \sin\beta \\
-\sin\beta & \cos\beta
\end{pmatrix}
\begin{pmatrix}
h_1 \\ h_2
\end{pmatrix}\;
\end{align} 
with
\begin{align}
m_H^2 = \frac{2\mu_{12}^2}{\sin 2\beta} \,, \qquad
m_h^2 = 2(\lambda_1 + \lambda_3) v^2 \;.
\label{mHh}
\end{align} 
At this point one should note that in the case of 2HDMs, the
combination $H^0 = (v_1h_1 + v_2h_2)/v$ has SM-like
couplings at the tree level.  But in general $H^0$ is not a physical
eigenstate.  In the limit where $H^0$ is aligned with one of the
physical $CP$-even scalars, is known as the \textit{alignment limit}
for 2HDMs.  \Eqn{eq:cpevn} shows that this is indeed the case in the
present model, viz., that the eigenstate $h$ is the same as $H^0$.
Thus the alignment limit emerges naturally~\cite{Dev:2014yca} in our
scenario.  Hence by identifying $h$ with the 125 GeV scalar observed
at the LHC, our model becomes consistent, by design, with the LHC
Higgs data~\cite{Khachatryan:2016vau}.

Looking at the spectrum, we can identify the following different
scenarios in regard to \Eqn{V2}.
\begin{enumerate}
\item If $\mu_1^2 = \mu_2^2$ and $\mu_{12}^2 = 0$, $V_2(\Phi)$ is
  completely $S_3$-symmetric.  In fact, the potential is invariant
  under a much bigger symmetry: an $\rm SO(2)$ symmetry under which
\begin{eqnarray}
\phi_1 &\to& \phi_1 \cos\alpha - \phi_2 \sin\alpha \,, \nonumber \\* 
\phi_2 &\to& \phi_1 \sin\alpha + \phi_2 \cos\alpha \,.
\end{eqnarray}
  Thus, after gauge symmetry breaking when the $\phi_i$'s develop
  vacuum expectation values (VEVs), we will have a massless scalar, a
  Goldstone boson as seen clearly from \Eqn{mHh}.  This is not the
  scenario that we advocate.

\item If $\mu_1^2 \neq \mu_2^2$ and $\mu_{12}^2 = 0$, the potential is
  not $S_3$ symmetric, but \Eqn{mHh} shows that we will still have a
  massless boson.  Thus, this is not our desired scenario either.

\item If $\mu_1^2 = \mu_2^2$ and $\mu_{12}^2 \neq 0$, there exists no
  massless scalar, but \Eqn{min} shows that we will now have
  $\tan\beta=1$ or $v_1=v_2$ because the potential has an exchange
  symmetry $\phi_1 \leftrightarrow \phi_2$.  As we discuss later, this
  scenario will be detrimental to our aim.

\item If $\mu_1^2 \neq \mu_2^2$ and $\mu_{12}^2 \neq 0$, there is no
  massless scalar and also $\tan\beta$ can be arbitrary.  This is the
  scenario that will be useful for us, implying that the soft-breaking
  terms are absolutely necessary.

\end{enumerate}

We now present the most general Yukawa couplings involving the $u_R$
quarks that is consistent with the gauge and $S_3$ symmetries.  The
$S_3$ symmetry cuts down on the number of Yukawa couplings
drastically, and we obtain only the following couplings involving
right-chiral $u$-type quarks:
\begin{eqnarray}
\mathscr L_Y^{(u)} &=&
\null - A_u \Big( 
\bar Q_1 \tilde\phi_1 + \bar Q_2\tilde\phi_2 \Big)
u_{3R} - B_u \Big\{ \Big( \bar Q_1\tilde\phi_2 + \bar
Q_2\tilde\phi_1\Big) u_{1R} + 
\Big( \bar Q_1\tilde\phi_1 -
\bar Q_2\tilde\phi_2 \Big)u_{2R} \Big\} \nonumber \\*
&& \null \qquad - C_u \bar Q_3 \Big( \tilde\phi_1 u_{1R} +
\tilde\phi_2u_{2R} \Big) 
 + {\rm h.c.}
\label{uYuk}
\end{eqnarray}
We have used the standard abbreviation $\tilde\phi_i = i \sigma_2
\phi_i^*$.  The Yukawa couplings of the $d_R$ quarks can be obtained
by replacing $u_{AR}$ by $d_{AR}$, $\{A,B,C\}_u$ by $\{A,B,C\}_d$, and
$\tilde\phi_i$ by $\phi_i$ in \Eqn{uYuk}.  Although the Yukawa
couplings, in general, may be complex, we will discuss later that all
but one phase can be absorbed in the field redefinitions.

After symmetry breaking, the mass matrices that arise in the quark
sector have the following form:
\begin{eqnarray}
{\cal M}_q = \frac v{\surd2}\begin{pmatrix}
B_q \sin\beta  &  B_q \cos\beta & A_q \cos\beta \\
B_q \cos\beta  & - B_q \sin\beta & A_q \sin\beta \\
C_q \cos\beta  & C_q \sin\beta & 0 \\
\end{pmatrix} \,,
\label{Mq}
\end{eqnarray}
where the subscripted index $q$ can take the value $u$ for
the up-type quarks, and $d$ for the down-type quarks.  It is
well-known that these matrices can be diagonalized through bi-unitary
transformations, e.g., one can find two unitary matrices $U_u$ and
$V_u$, for the up-sector, such that $U_u {\cal M}_u V_u^\dagger$ is
diagonal.  The CKM matrix is then given by $U_u U_d^\dagger$.

The matrices $U_u$ and $U_d$ are the unitary matrices which
diagonalize, through similarity transformations, the hermitian
matrices ${\cal M}_u {\cal M}_u^\dagger$ and ${\cal M}_d {\cal
  M}_d^\dagger$ respectively.  From \Eqn{Mq}, we obtain
\begin{eqnarray}
{\cal M}_q {\cal M}_q^\dagger = 
\frac12 v^2 
\begin{pmatrix}
a_q^2 \cos^2\beta + b_q^2 & \frac12 a_q^2 \sin 2\beta  & B_qC_q^* \sin
2\beta \\ 
\frac12 a_q^2  \sin 2\beta  & a_q^2 \sin^2\beta + b_q^2  
& B_qC_q^* \cos2\beta \\ 
B_q^*C_q  \sin 2\beta & B_q^*C_q \cos2\beta & c_q^2 \\ 
\end{pmatrix} \,,
\label{MMdagABC}
\end{eqnarray}
where $a_q = |A_q|$ etc.  Clearly, the three eigenvalues of ${\cal
  M}_u {\cal M}_u^\dagger$ would be the mass squared of the three up
sector quarks, namely $m_u^2,m_c^2$ and $m_t^2$, and the three
eigenvalues of ${\cal M}_d {\cal M}_d^\dagger$ would be $m_d^2$,
$m_s^2$ and $m_b^2$.  The eigenvalues can be obtained by solving the
characteristic equation of the general matrix in \Eqn{MMdagABC}.
Introducing the shorthand notation
\begin{equation}
x = {2m^2 \over v^2}
\label{x}
\end{equation}
for any fermion with mass $m$, this characteristic equation has the
following form:
\begin{eqnarray}
x^3 - (a^2 + 2b^2 + c^2) x^2 + (a^2 + b^2)
      (b^2+c^2) x - a^2b^2c^2 \sin^2 3\beta = 0 \,,
\label{charEq}
\end{eqnarray}
with subscripts $u$ or $d$ attached to the Yukawa couplings, as the
case may be.   Note that this equation is free from the phase of
$BC^*$, which is the only phase that is present in \Eqn{MMdagABC}.

Looking at the Lagrangian of \Eqn{uYuk} and the corresponding
Lagrangian involving $d_{iR}$, we see why only one phase is present in
\Eqn{MMdagABC}.  Any phase of $A_u$ and $A_d$ can be absorbed by
redefining the fields $u_{3R}$ and $d_{3R}$.  After this, both $B_u$
and $B_d$ can be made real by redefining the fields $u_{1R}, u_{2R}$
and $d_{1R}, d_{2R}$.  Finally, either $C_u$ or $C_d$ can be made real
by choosing the phase of $Q_3$, but one of them remains complex.
Alternatively, one can make both $C_u$ or $C_d$ real first, by
redefining the right-chiral quark fields, and then either $B_u$ or
$B_d$ can be made real by choosing the phases of $Q_1$ and $Q_2$.
Either way, one of the $C_q$'s or one of the $B_q$'s can be complex in
the most general case.  In what follows, we will assume that all
Yukawa couplings are real, and use the lower-case symbols for them.

Before entering into a discussion of the eigenvalues obtained as
solutions of \Eqn{charEq}, let us have some idea of the form of the
diagonalizing matrix.  As a first step, we can diagonalize only the
terms in \Eqn{MMdagABC} that are proportional to $a_q^2$.  This is
done, e.g., by a matrix
\begin{eqnarray}
{\cal U} = \begin{pmatrix}
0 & 0 & 1 \\ 
\sin\beta & -\cos\beta & 0 \\
\cos\beta & \sin\beta & 0 \\ 
\end{pmatrix} \,.
\end{eqnarray}
Note that this matrix does not depend on the Yukawa couplings, and is
therefore the same for the up-type and down-type mass matrices.  
Applying a similarity transformation with this matrix on ${\cal
  MM}^\dagger$, we obtain
\begin{eqnarray}
M^2 = {\cal UMM^\dagger U^\dagger} = \frac12 v^2
\begin{pmatrix}
c^2 & -bc \cos3\beta & bc \sin3\beta \\ 
- bc \cos3\beta & b^2 & 0 \\
 bc \sin3\beta & 0 & a^2 + b^2 
\end{pmatrix} \,,
\label{UMMU}
\end{eqnarray}
with subscripts $u$ and $d$ attached for quarks of positive and
negative charges respectively.

In the preamble of the article, we said that we want to relate the
almost-masslessness of first generation quarks with the
almost-block-diagonal form of the CKM matrix.  We now narrow down
the scenario in which we can have one zero eigenvalue in both
up-type and down-type quark sector, as well as a block-diagonal CKM
matrix.

First we note that if one solution of \Eqn{charEq} is zero, then the
$x$-independent term should vanish in that equation.  In this case,
the eigenvalues of ${\cal MM}^\dagger$ are given by
\begin{eqnarray}
0, \qquad \frac12 v^2 (b^2+c^2), \qquad \frac12 v^2 (a^2+b^2) \,.
\label{mass^2}
\end{eqnarray}
For the diagonalizing matrix, we now consider two different
cases.  

\subsubsection*{Case 1: Some Yukawa couplings vanish}
Surely, the $x$-independent term in \Eqn{charEq} can vanish if at
least one of the Yukawa couplings is zero.  Looking at \Eqn{UMMU}, we
see that $a=0$ does not make $M^2$ block-diagonal, so we reject this
possibility.  If either $b$ or $c$ vanishes, the matrix $M^2$
becomes completely diagonal.  This means that for $b$ or $c = 0$, the
same matrix $\cal U$ will diagonalize both ${\cal M}_u {\cal
  M}_u^\dagger$ and ${\cal M}_d {\cal M}_d^\dagger$ making the CKM
matrix a unit matrix.  Therefore making some Yukawa coupling vanish to
obtain one zero mass does not produce the desirable block-diagonal
structure of the CKM matrix.

One should recall that making $\mu_1^2 = \mu_2^2$ in
\Eqn{V2} had led to $\tan\beta = 1$, which in view of
\Eqn{charEq} demands that one of the Yukawas must be zero in order to
obtain zero mass eigenvalue.  For this reason we discard this
particular form of $V_2(\Phi)$.

\subsubsection*{Case 2: $\sin3\beta=0$}
However, there is a second and more attractive possibility.  From the
characteristic equation, \Eqn{charEq}, one can see that zero
eigenvalue can also be ensured if
\begin{eqnarray}
\sin 3\beta = 0 \,.
\label{sin3beta}
\end{eqnarray}
Discarding the trivial solution $\beta=0$, we obtain the solution
$\beta = \frac13\pi$ which implies that $\tan\beta = \surd3$ i.e.,
$v_2 = \surd3v_1 = \surd3v/2$.\footnote{While this VEV alignment is
  useful for our consequent discussions we note that in case of a
  three Higgs doublet model with $S_3$ symmetry, the minimisation of
  potential leads to a vev alignment $v_1 = \surd3v_2$ and such
  alignment implies a residual $\mathbb{Z}_2$
  symmetry~\cite{Das:2014fea}.  In the present case, however, no such
  implications are possible.}
Looking at \Eqn{UMMU} now, we see that this value of $\beta$ also
makes the matrix $M^2$ block-diagonal, and one obtains
\begin{eqnarray}
M^2 = {\cal UMM^\dagger U^\dagger} = \frac12 v^2
\begin{pmatrix}
c^2 & bc & 0 \\ 
bc & b^2 & 0 \\
0 & 0 & a^2 + b^2 
\end{pmatrix} \,.
\end{eqnarray}
Notice that the third generation has been singled out, and therefore
$v\sqrt{(a^2+b^2)/2}$ can be readily identified with the mass of the
third generation quark.  In order that it be much heavier than the
quarks in the first two generations, we need
\begin{eqnarray}
a^2 \gg b^2, c^2 
\label{gg}
\end{eqnarray}
in both up and down sectors.

Complete diagonalization would require a further similarity
transformation affecting the upper $2 \times 2$ block.  This will
involve the values of the Yukawa couplings.  Thus, we obtain
\begin{eqnarray}
U_u = {\cal O}_u {\cal U} \,, \qquad
U_d = {\cal O}_d {\cal U} \,, 
\label{UuUd}
\end{eqnarray}
where 
\begin{eqnarray}
{\cal O}_q = \begin{pmatrix}
\cos \theta_q & - \sin\theta_q & 0 \\
 \sin\theta_q & \cos\theta_q & 0 \\ 
0 & 0 & 1
\end{pmatrix} \,,
\end{eqnarray}
with
\begin{eqnarray}
\tan \theta_q = {c_q \over b_q} \,.
\label{b/c}
\end{eqnarray}
From \Eqn{UuUd} the CKM matrix can now be written as,
\begin{eqnarray}
V_{\rm CKM} = U_u U_d^\dagger = {\cal O}_u {\cal O}_d^\dagger = 
\begin{pmatrix}
\cos(\theta_u - \theta_d) & - \sin(\theta_u - \theta_d) & 0 \\
 \sin(\theta_u - \theta_d) & \cos(\theta_u - \theta_d) & 0 \\ 
0 & 0 & 1
\end{pmatrix} \,.
\end{eqnarray}
Thus the difference $\theta_u-\theta_d$, which can be identified with
the Cabibbo angle, $\theta_{C}$.

In passing, we make a point about the VEV alignment, i.e., the value
of $\beta$, dictated by \Eqn{sin3beta}.  It reflects our choice of the
representation for $S_3$.  Had we chosen a different representation,
the value of $\beta$ would in general be different.  But the physical
implications should be independent of the representation, and so the
block-diagonal form of the CKM matrix would still result.

Having reproduced the leading order effects of the mixing matrix in
the Wolfenstein parametrization as a consequence of the masslessness
of the first generation quarks, we now explore whether one can do
better.  So far, the conclusions that we derived came from
\Eqn{sin3beta}, which is a statement about the relative magnitude of
the VEVs of the two Higgs doublets.  Note that this relation is not
protected by any symmetry.  
Suppose we deviate from \Eqn{sin3beta} by a small amount such that
\begin{eqnarray}
\sin 3\beta = \delta \,.
\label{delta}
\end{eqnarray}
Since $\delta$ is expected to be small, we do not expect the heavier
quark masses to be altered very much by this change.  The sums of
eigenvalues etc.  will also not change appreciably.  The only thing that
will change dramatically is the product of all eigenvalues, which
should be the $x$-independent term in \Eqn{charEq}.  Therefore the
first generation quark masses will be given, in the notation of
\Eqn{x}, by
\begin{subequations}
\begin{align}
x_u = {2m_u^2 \over v^2} & \approx {a_u^2b_u^2c_u^2 \delta^2 \over
  (a_u^2+b_u^2) (b_u^2+c_u^2)} 
\approx  {b_u^2c_u^2 \delta^2 \over b_u^2+c_u^2} \\ 
x_d = {2m_d^2 \over v^2} & \approx {a_d^2b_d^2c_d^2 \delta^2 \over
  (a_d^2+b_d^2) (b_d^2+c_d^2)} 
\approx  {b_d^2c_d^2 \delta^2 \over b_d^2+c_d^2}  \,, 
\end{align}
\end{subequations}
where in the last step we have used the hierarchy mentioned in
\Eqn{gg}.  Now, using \Eqn{b/c}, we can write
\begin{subequations}
\begin{eqnarray}
\label{eq:musq}
m_u^2 &\approx& \frac12 v^2 \delta^2 \times (b_u^2+c_u^2)
\sin^2\theta_u \cos^2 \theta_u \approx \frac14 m_c^2 \delta^2 \sin^2
2\theta_u \\* 
\label{eq:mdsq}
m_d^2 &\approx& \frac12 v^2 \delta^2 \times  (b_d^2+c_d^2)
\sin^2\theta_d \cos^2 \theta_d \approx \frac14 m_s^2 \delta^2 \sin^2
2\theta_d \,.
\end{eqnarray}
\end{subequations}
Since $\theta_u  - \theta_d  = \theta_{C}$, \Eqs{eq:musq}{eq:mdsq}
can be solved for $\delta$ and $\theta_u$ or $\theta_d $.  Taking
all the uncertainties into account we have found $\delta > 0.2$ which
is inconsistent with our assumption of small $\delta$ in
\Eqn{delta}.  Therefore, this minimal framework is not sufficient to
reproduce the observed masses of the first generation quarks.

Now, for completeness, we comment on the flavour changing neutral
currents (FCNC) in our model.  To set up the notations we first lay
out the Yukawa Lagrangian for 2HDM in the following form:
\begin{align}
\label{eq:l2hdmyuk}
\mathscr{L}_Y = - \sum_{k =
  1}^2\left[\bar{\mathbf Q}_L\Gamma_k \phi_k \mathbf d_{R} +
  \bar{\mathbf Q}_L\Delta_k \tilde{\phi}_k \mathbf u_{R}\right] +
        {\rm h.c.},  
\end{align}
where we have kept the notation for the field the same as in
\Eqn{uYuk} but put them in boldface font to remind ourselves that the
generation indices have been suppressed.  Unlike Eq.  (\ref{uYuk}),
here we also take into account the Yukawa Lagrangian for the down
sector too.  Here $\Delta_{1,2}$ and $\Gamma_{1,2}$ represent the
Yukawa matrices in the up and down sectors respectively.  By comparing
\Eqs{eq:l2hdmyuk}{uYuk} we can write,
\begin{eqnarray}
\Delta_1 = 
\begin{pmatrix}
0 & b_u & a_u \\
b_u & 0 & 0\\
c_u & 0 & 0
\end{pmatrix} \,, \qquad 
\Delta_2 = 
\begin{pmatrix}
b_u & 0 & 0 \\
0  &  -b_u & a_u \\
0  &   c_u & 0
\end{pmatrix} \,,
\end{eqnarray}
and the $\Gamma_k$'s can be obtained by replacing the subscript $u$ by
the subscript $d$ in the matrices.  Now, the Yukawa Lagrangian in
terms of physical fields can be written as
\begin{eqnarray}
\mathscr{L}_{\rm Yuk} &=& -\frac{h}{v}(\bar{\mathbf d} D_d \mathbf d + 
              \bar{\mathbf u}D_u \mathbf u) + 
              \frac{H}{v} \left[ \bar{\mathbf d}\left(
                N_d P_{R} + N_d^\dagger P_L
                \right) \mathbf d + \bar{\mathbf u}\left(
                N_u P_{R} + N_u^\dagger P_L
                \right) \mathbf u
                \right] \notag \\
                &&-\frac{iA}{v}\left[\bar{\mathbf u}\left(
                N_u P_{R} - N_u^\dagger P_L
                \right) \mathbf u - \bar{\mathbf d}\left(
                N_d P_{R} - N_d^\dagger P_L
                \right) \mathbf d
                \right] \notag \\
                &&+ \frac{\surd2H^+}{v}\bar{\mathbf u}\left[
                V_{\rm CKM}N_d P_{R} - 
                N_u^\dagger V_{\rm CKM}P_L\right] \mathbf d + {\rm h.c.},
\label{eq:LYukfcnc}                
\end{eqnarray}
where $D_u$ and $D_d$ are the diagonal mass matrices in the up and
down sectors respectively.  Note that the SM-like scalar, $h$, does not
have any FCNC couplings.  This is a direct consequence of the natural
alignment that we have talked about earlier.  The matrices $N_u$ and
$N_d$, in \Eqn{eq:LYukfcnc}, carry the information of FCNC in the up
and down sectors respectively and are given by,
\begin{subequations}
\label{NuNd}
\begin{align}
\label{Nu}
N_u  &= \frac1{\surd2}
              U_u (\Delta_1v_2 - \Delta_2v_1) V_u^\dagger, \\
\label{Nd}
N_d  &= \frac1{\surd2}
         U_d (\Gamma_1v_2 - \Gamma_2v_1) V_d^\dagger \,.
\end{align}
\end{subequations}
Note that the expressions for $V_u$ and $V_d$ can be obtained from
diagonalizing $\mathcal{M^\dagger M}$ for both up and down sectors.
The matrices $\mathcal{M^\dagger M}$ can be obtained from
$\mathcal{MM^\dagger}$ by making the interchange $a \leftrightarrow c$
in the Yukawa couplings.  Because of this interchange, the matrix $V$
is different from $U$ in two respects.  First, the matrix
corresponding to $\cal U$ should have the last two rows interchanged
so that the eigenvalues can occur in the same order.  Second, the
angle $\theta_q$ should be replaced by $\theta'_q$, which will be
given by
\begin{eqnarray}
\tan \theta'_q = {a_q \over b_q} \,.
\end{eqnarray}
In view of the hierarchy mentioned in \Eqn{gg}, we can use these to write 
\begin{eqnarray}
\sin \theta'_q \approx 1 \,, \qquad \cos \theta'_q \approx {b_q \over
  a_q} \,,
\end{eqnarray}
neglecting higher order terms in $b_q/a_q$.  Replacing the Yukawa
couplings by the mass eigenvalues and the angles $\theta_q$, we obtain 
\begin{eqnarray}
N_u  \approx \begin{pmatrix}
-\frac32 m_c\sin 2\theta_u  & 0 & - m_t \sin \theta_u  \\
\frac12 m_c(3\cos^2\theta_u - 1 ) & 0 & m_t\cos \theta_u  \\
0 & m_c\cos \theta_u  & 0 
\end{pmatrix} \,,
\label{Numat}
\end{eqnarray}
neglecting corrections of order $m_c/m_t$.  
A similar expression for $N_d$ can be obtained from \Eqn{Numat}
by replacing $\theta_u , m_c, m_t$ with $\theta_d , m_s,
m_b$ respectively.  Thus the FCNCs are uniquely determined by
$\theta_u$ or $\theta_d$.  One should keep in mind that this
represents the FCNC couplings at the leading order, i.e., when the CKM
matrix is block-diagonal.  In a more complete framework where the CKM
matrix can be reproduced exactly these FCNC matrices are expected to
get small corrections.

A trivial but viable solution to the FCNC problem would be to make all
the scalars except $h$ sufficiently heavy.  Moreover, the bounds from
the electroweak $T$-parameter can also be evaded if the non-standard
scalars, $H, A$ and $H^{\pm}$ are nearly
degenerate~\cite{Grimus:2007if, Bhattacharyya:2013rya}.

In summary, we connect two apparently disjoint experimental
observations namely, the tiny masses of first generation of quarks and
the near block-diagonal structure of the CKM matrix in a simple set-up
of 2HDM with an $S_3$ symmetry.  We attribute these two features of
the quark sector to Nature's choice of a particular value of
$\tan\beta$.  An added bonus of our model is the existence of a light
scalar, which can be identified with the 125 GeV Higgs observed at the
LHC, in view of a naturally emerging alignment limit.  Admittedly, the
exact CKM matrix and correct non-zero masses for the first generation
of quarks could not be reproduced in this minimalistic scenario.
Perhaps our set-up can be taken as a constituent towards a more
elaborate framework which can address the full quark structure.

\paragraph*{Acknowledgements\,:}
DD thanks Arcadi Santamaria and Anirban Kundu for useful discussions.  The work of UKD is
supported by Department of Science and Technology, Government of India
under the fellowship reference number PDF/2016/001087 (SERB National
Post-Doctoral fellowship).


\bibliographystyle{JHEP} 
\bibliography{s3_2hdm.bib}

\providecommand{\href}[2]{#2}\begingroup\raggedright\begin{thebibliography}{10}

\bibitem{Altarelli:2010gt}
G.~Altarelli and F.~Feruglio, {\it {Discrete Flavor Symmetries and Models of
  Neutrino Mixing}},  {\em Rev. Mod. Phys.} {\bf 82} (2010) 2701--2729,
  [\href{http://arxiv.org/abs/1002.0211}{{\tt arXiv:1002.0211}}].

\bibitem{Ishimori:2010au}
H.~Ishimori, T.~Kobayashi, H.~Ohki, Y.~Shimizu, H.~Okada, and M.~Tanimoto, {\it
  {Non-Abelian Discrete Symmetries in Particle Physics}},  {\em Prog. Theor.
  Phys. Suppl.} {\bf 183} (2010) 1--163,
  [\href{http://arxiv.org/abs/1003.3552}{{\tt arXiv:1003.3552}}].

\bibitem{Wolfenstein:1983yz}
L.~Wolfenstein, {\it {Parametrization of the Kobayashi-Maskawa Matrix}},  {\em
  Phys. Rev. Lett.} {\bf 51} (1983) 1945.

\bibitem{Pakvasa:1977in}
S.~Pakvasa and H.~Sugawara, {\it {Discrete Symmetry and Cabibbo Angle}},  {\em
  Phys. Lett.} {\bf B73} (1978) 61--64.

\bibitem{Koide:1999mx}
Y.~Koide, {\it {Universal seesaw mass matrix model with an S(3) symmetry}},
  {\em Phys.Rev.} {\bf D60} (1999) 077301,
  [\href{http://arxiv.org/abs/hep-ph/9905416}{{\tt hep-ph/9905416}}].

\bibitem{Kubo:2003iw}
J.~Kubo, A.~Mondragon, M.~Mondragon, and E.~Rodriguez-Jauregui, {\it {The
  Flavor symmetry}},  {\em Prog. Theor. Phys.} {\bf 109} (2003) 795--807,
  [\href{http://arxiv.org/abs/hep-ph/0302196}{{\tt hep-ph/0302196}}]. [Erratum:
  Prog. Theor. Phys.114,287(2005)].

\bibitem{Harrison:2003aw}
P.~Harrison and W.~Scott, {\it {Permutation symmetry, tri - bimaximal neutrino
  mixing and the S3 group characters}},  {\em Phys.Lett.} {\bf B557} (2003) 76,
  [\href{http://arxiv.org/abs/hep-ph/0302025}{{\tt hep-ph/0302025}}].

\bibitem{Chen:2004rr}
S.-L. Chen, M.~Frigerio, and E.~Ma, {\it {Large neutrino mixing and normal mass
  hierarchy: A Discrete understanding}},  {\em Phys. Rev.} {\bf D70} (2004)
  073008, [\href{http://arxiv.org/abs/hep-ph/0404084}{{\tt hep-ph/0404084}}].
  [Erratum: Phys. Rev.D70,079905(2004)].

\bibitem{Koide:2006vs}
Y.~Koide, {\it {S(3) symmetry and neutrino masses and mixings}},  {\em Eur.
  Phys. J.} {\bf C50} (2007) 809--816,
  [\href{http://arxiv.org/abs/hep-ph/0612058}{{\tt hep-ph/0612058}}].

\bibitem{Mondragon:2007af}
A.~Mondragon, M.~Mondragon, and E.~Peinado, {\it {Lepton masses, mixings and
  FCNC in a minimal S(3)-invariant extension of the Standard Model}},  {\em
  Phys. Rev.} {\bf D76} (2007) 076003,
  [\href{http://arxiv.org/abs/0706.0354}{{\tt arXiv:0706.0354}}].

\bibitem{Chen:2007zj}
C.-Y. Chen and L.~Wolfenstein, {\it {Consequences of approximate S(3) symmetry
  of the neutrino mass matrix}},  {\em Phys. Rev.} {\bf D77} (2008) 093009,
  [\href{http://arxiv.org/abs/0709.3767}{{\tt arXiv:0709.3767}}].

\bibitem{Jora:2009gz}
R.~Jora, J.~Schechter, and M.~Naeem~Shahid, {\it {Perturbed S(3) neutrinos}},
  {\em Phys.Rev.} {\bf D80} (2009) 093007,
  [\href{http://arxiv.org/abs/0909.4414}{{\tt arXiv:0909.4414}}].

\bibitem{Kaneko:2010rx}
T.~Kaneko and H.~Sugawara, {\it {Broken $S_3$ Symmetry in Flavor Physics}},
  {\em Phys. Lett.} {\bf B697} (2011) 329--332,
  [\href{http://arxiv.org/abs/1011.5748}{{\tt arXiv:1011.5748}}].

\bibitem{Teshima:2011wg}
T.~Teshima and Y.~Okumura, {\it {Quark/lepton mass and mixing in $S_3$
  invariant model and CP-violation of neutrino}},  {\em Phys. Rev.} {\bf D84}
  (2011) 016003, [\href{http://arxiv.org/abs/1103.6127}{{\tt
  arXiv:1103.6127}}].

\bibitem{Dev:2011qy}
S.~Dev, S.~Gupta, and R.~R. Gautam, {\it {Broken $S_3$ Symmetry in the Neutrino
  Mass Matrix}},  {\em Phys.Lett.} {\bf B702} (2011) 28--33,
  [\href{http://arxiv.org/abs/1106.3873}{{\tt arXiv:1106.3873}}].

\bibitem{Dias:2012bh}
A.~G. Dias, A.~C.~B. Machado, and C.~C. Nishi, {\it {An $S_3$ Model for Lepton
  Mass Matrices with Nearly Minimal Texture}},  {\em Phys. Rev.} {\bf D86}
  (2012) 093005, [\href{http://arxiv.org/abs/1206.6362}{{\tt
  arXiv:1206.6362}}].

\bibitem{Meloni:2012ci}
D.~Meloni, {\it {$S_3$ as a flavour symmetry for quarks and leptons after the
  Daya Bay result on $\theta_{13}$}},  {\em JHEP} {\bf 05} (2012) 124,
  [\href{http://arxiv.org/abs/1203.3126}{{\tt arXiv:1203.3126}}].

\bibitem{Dev:2012ns}
S.~Dev, R.~R. Gautam, and L.~Singh, {\it {Broken $S_3$ Symmetry in the Neutrino
  Mass Matrix and Non-Zero $\theta_{13}$}},  {\em Phys. Lett.} {\bf B708}
  (2012) 284--289, [\href{http://arxiv.org/abs/1201.3755}{{\tt
  arXiv:1201.3755}}].

\bibitem{Zhou:2011nu}
S.~Zhou, {\it {Relatively large theta13 and nearly maximal theta23 from the
  approximate S3 symmetry of lepton mass matrices}},  {\em Phys. Lett.} {\bf
  B704} (2011) 291--295, [\href{http://arxiv.org/abs/1106.4808}{{\tt
  arXiv:1106.4808}}].

\bibitem{Ma:2013zca}
E.~Ma and B.~Melic, {\it {Updated $S_{3}$ model of quarks}},  {\em Phys. Lett.}
  {\bf B725} (2013) 402--406, [\href{http://arxiv.org/abs/1303.6928}{{\tt
  arXiv:1303.6928}}].

\bibitem{Benaoum:2013ji}
H.~B. Benaoum, {\it {Broken $S_3$ Neutrinos}},  {\em Phys. Rev.} {\bf D87}
  (2013) 073010, [\href{http://arxiv.org/abs/1302.0950}{{\tt
  arXiv:1302.0950}}].

\bibitem{Ma:2014qra}
E.~Ma and R.~Srivastava, {\it {Dirac or inverse seesaw neutrino masses with
  $B-L$ gauge symmetry and $S_3$ flavor symmetry}},  {\em Phys. Lett.} {\bf
  B741} (2015) 217--222, [\href{http://arxiv.org/abs/1411.5042}{{\tt
  arXiv:1411.5042}}].

\bibitem{Das:2015sca}
D.~Das, U.~K. Dey, and P.~B. Pal, {\it {$S_3$ symmetry and the quark mixing
  matrix}},  {\em Phys. Lett.} {\bf B753} (2016) 315--318,
  [\href{http://arxiv.org/abs/1507.06509}{{\tt arXiv:1507.06509}}].

\bibitem{Cogollo:2016dsd}
D.~Cogollo and J.~P. Silva, {\it {Two Higgs doublet models with an $S_3$
  symmetry}},  {\em Phys. Rev.} {\bf D93} (2016), no.~9 095024,
  [\href{http://arxiv.org/abs/1601.02659}{{\tt arXiv:1601.02659}}].

\bibitem{Pramanick:2016mdp}
S.~Pramanick and A.~Raychaudhuri, {\it {Neutrino mass model with $S_3$ symmetry
  and seesaw interplay}},  {\em Phys. Rev.} {\bf D94} (2016), no.~11 115028,
  [\href{http://arxiv.org/abs/1609.06103}{{\tt arXiv:1609.06103}}].

\bibitem{Varzielas:2016zuo}
I.~de~Medeiros~Varzielas, R.~W. Rasmussen, and J.~Talbert, {\it {Bottom-Up
  Discrete Symmetries for Cabibbo Mixing}},  {\em Int. J. Mod. Phys.} {\bf A32}
  (2017), no.~06n07 1750047, [\href{http://arxiv.org/abs/1605.03581}{{\tt
  arXiv:1605.03581}}].

\bibitem{Deshpande:1991zh}
N.~G. Deshpande, M.~Gupta, and P.~B. Pal, {\it {Flavor changing processes and
  CP violation in $S(3) \times Z(3)$ model}},  {\em Phys. Rev.} {\bf D45}
  (1992) 953--957.

\bibitem{Kubo:2004ps}
J.~Kubo, H.~Okada, and F.~Sakamaki, {\it {Higgs potential in minimal S(3)
  invariant extension of the standard model}},  {\em Phys. Rev.} {\bf D70}
  (2004) 036007, [\href{http://arxiv.org/abs/hep-ph/0402089}{{\tt
  hep-ph/0402089}}].

\bibitem{Teshima:2005bk}
T.~Teshima, {\it {Flavor mass and mixing and S(3) symmetry: An S(3) invariant
  model reasonable to all}},  {\em Phys. Rev.} {\bf D73} (2006) 045019,
  [\href{http://arxiv.org/abs/hep-ph/0509094}{{\tt hep-ph/0509094}}].

\bibitem{Koide:2005ep}
Y.~Koide, {\it {Permutation symmetry S(3) and VEV structure of flavor-triplet
  Higgs scalars}},  {\em Phys. Rev.} {\bf D73} (2006) 057901,
  [\href{http://arxiv.org/abs/hep-ph/0509214}{{\tt hep-ph/0509214}}].

\bibitem{Beltran:2009zz}
O.~F. Beltran, M.~Mondragon, and E.~Rodriguez-Jauregui, {\it {Conditions for
  vacuum stability in an S(3) extension of the standard model}},  {\em
  J.Phys.Conf.Ser.} {\bf 171} (2009) 012028.

\bibitem{Ferreira:2010hy}
P.~M. Ferreira, M.~Maniatis, O.~Nachtmann, and J.~P. Silva, {\it {CP properties
  of symmetry-constrained two-Higgs-doublet models}},  {\em JHEP} {\bf 08}
  (2010) 125, [\href{http://arxiv.org/abs/1004.3207}{{\tt arXiv:1004.3207}}].

\bibitem{Bhattacharyya:2010hp}
G.~Bhattacharyya, P.~Leser, and H.~Pas, {\it {Exotic Higgs boson decay modes as
  a harbinger of $S_3$ flavor symmetry}},  {\em Phys. Rev.} {\bf D83} (2011)
  011701, [\href{http://arxiv.org/abs/1006.5597}{{\tt arXiv:1006.5597}}].

\bibitem{Bhattacharyya:2012ze}
G.~Bhattacharyya, P.~Leser, and H.~Pas, {\it {Novel signatures of the Higgs
  sector from S3 flavor symmetry}},  {\em Phys. Rev.} {\bf D86} (2012) 036009,
  [\href{http://arxiv.org/abs/1206.4202}{{\tt arXiv:1206.4202}}].

\bibitem{Teshima:2012cg}
T.~Teshima, {\it {Higgs potential in $S_3$ invariant model for quark/lepton
  mass and mixing}},  {\em Phys. Rev.} {\bf D85} (2012) 105013,
  [\href{http://arxiv.org/abs/1202.4528}{{\tt arXiv:1202.4528}}].

\bibitem{Machado:2012ed}
A.~C.~B. Machado and V.~Pleitez, {\it {A model with two inert scalar
  doublets}},  {\em Annals Phys.} {\bf 364} (2016) 53--67,
  [\href{http://arxiv.org/abs/1205.0995}{{\tt arXiv:1205.0995}}].

\bibitem{Barradas-Guevara:2014yoa}
E.~Barradas-Guevara, O.~Félix-Beltrán, and E.~Rodríguez-Jáuregui, {\it
  {Trilinear self-couplings in an S(3) flavored Higgs model}},  {\em Phys.
  Rev.} {\bf D90} (2014), no.~9 095001,
  [\href{http://arxiv.org/abs/1402.2244}{{\tt arXiv:1402.2244}}].

\bibitem{Das:2014fea}
D.~Das and U.~K. Dey, {\it {Analysis of an extended scalar sector with $S_3$
  symmetry}},  {\em Phys. Rev.} {\bf D89} (2014), no.~9 095025,
  [\href{http://arxiv.org/abs/1404.2491}{{\tt arXiv:1404.2491}}]. [Erratum:
  Phys. Rev.D91,no.3,039905(2015)].

\bibitem{Barradas-Guevara:2015rea}
E.~Barradas-Guevara, O.~Félix-Beltrán, and E.~Rodríguez-Jáuregui, {\it {CP
  breaking in $S(3)$ flavoured Higgs model}},
  \href{http://arxiv.org/abs/1507.05180}{{\tt arXiv:1507.05180}}.

\bibitem{Merchand:2016ldu}
M.~Merchand and M.~Sher, {\it {Three doublet lepton-specific model}},  {\em
  Phys. Rev.} {\bf D95} (2017), no.~5 055004,
  [\href{http://arxiv.org/abs/1611.06887}{{\tt arXiv:1611.06887}}].

\bibitem{Emmanuel-Costa:2016vej}
D.~Emmanuel-Costa, O.~M. Ogreid, P.~Osland, and M.~N. Rebelo, {\it {Spontaneous
  symmetry breaking in the $S_3$-symmetric scalar sector}},  {\em JHEP} {\bf
  02} (2016) 154, [\href{http://arxiv.org/abs/1601.04654}{{\tt
  arXiv:1601.04654}}]. [Erratum: JHEP08,169(2016)].

\bibitem{Xing:2010iu}
Z.-z. Xing, D.~Yang, and S.~Zhou, {\it {Broken $S_3$ Flavor Symmetry of Leptons
  and Quarks: Mass Spectra and Flavor Mixing Patterns}},  {\em Phys. Lett.}
  {\bf B690} (2010) 304--310, [\href{http://arxiv.org/abs/1004.4234}{{\tt
  arXiv:1004.4234}}].

\bibitem{Canales:2012dr}
F.~Gonzalez~Canales, A.~Mondragon, and M.~Mondragon, {\it {The $S_3$ Flavour
  Symmetry: Neutrino Masses and Mixings}},  {\em Fortsch. Phys.} {\bf 61}
  (2013) 546--570, [\href{http://arxiv.org/abs/1205.4755}{{\tt
  arXiv:1205.4755}}].

\bibitem{Canales:2013cga}
F.~González~Canales, A.~Mondragón, M.~Mondragón, U.~J. Saldaña~Salazar, and
  L.~Velasco-Sevilla, {\it {Quark sector of S3 models: classification and
  comparison with experimental data}},  {\em Phys. Rev.} {\bf D88} (2013)
  096004, [\href{http://arxiv.org/abs/1304.6644}{{\tt arXiv:1304.6644}}].

\bibitem{Hernandez:2014vta}
A.~E. Cárcamo~Hernández, R.~Martinez, and J.~Nisperuza, {\it {$S_3$ discrete
  group as a source of the quark mass and mixing pattern in $331$ models}},
  {\em Eur. Phys. J.} {\bf C75} (2015), no.~2 72,
  [\href{http://arxiv.org/abs/1401.0937}{{\tt arXiv:1401.0937}}].

\bibitem{Hernandez:2015dga}
A.~E. Cárcamo~Hernández, I.~de~Medeiros~Varzielas, and E.~Schumacher, {\it
  {Fermion and scalar phenomenology of a two-Higgs-doublet model with $S_3$}},
  {\em Phys. Rev.} {\bf D93} (2016), no.~1 016003,
  [\href{http://arxiv.org/abs/1509.02083}{{\tt arXiv:1509.02083}}].

\bibitem{Hernandez:2015zeh}
A.~E. Cárcamo~Hernández, I.~de~Medeiros~Varzielas, and N.~A. Neill, {\it
  {Novel Randall-Sundrum model with $S_{3}$ flavor symmetry}},  {\em Phys.
  Rev.} {\bf D94} (2016), no.~3 033011,
  [\href{http://arxiv.org/abs/1511.07420}{{\tt arXiv:1511.07420}}].

\bibitem{Gomez-Izquierdo:2017rxi}
J.~C. Gómez-Izquierdo, {\it {Non-Minimal Flavored ${\bf S}_{3}\otimes {\bf
  Z}_{2} $ Left-Right Symmetric Model}},
  \href{http://arxiv.org/abs/1701.01747}{{\tt arXiv:1701.01747}}.

\bibitem{Branco:2011iw}
G.~C. Branco, P.~M. Ferreira, L.~Lavoura, M.~N. Rebelo, M.~Sher, and J.~P.
  Silva, {\it {Theory and phenomenology of two-Higgs-doublet models}},  {\em
  Phys. Rept.} {\bf 516} (2012) 1--102,
  [\href{http://arxiv.org/abs/1106.0034}{{\tt arXiv:1106.0034}}].

\bibitem{Khachatryan:2016vau}
{\bf ATLAS, CMS} Collaboration, G.~Aad et~al., {\it {Measurements of the Higgs
  boson production and decay rates and constraints on its couplings from a
  combined ATLAS and CMS analysis of the LHC pp collision data at $ \sqrt{s}=7
  $ and 8 TeV}},  {\em JHEP} {\bf 08} (2016) 045,
  [\href{http://arxiv.org/abs/1606.02266}{{\tt arXiv:1606.02266}}].

\bibitem{Dev:2014yca}
P.~S. Bhupal~Dev and A.~Pilaftsis, {\it {Maximally Symmetric Two Higgs Doublet
  Model with Natural Standard Model Alignment}},  {\em JHEP} {\bf 12} (2014)
  024, [\href{http://arxiv.org/abs/1408.3405}{{\tt arXiv:1408.3405}}].
  [Erratum: JHEP11,147(2015)].

\bibitem{Grimus:2007if}
W.~Grimus, L.~Lavoura, O.~M. Ogreid, and P.~Osland, {\it {A Precision
  constraint on multi-Higgs-doublet models}},  {\em J. Phys.} {\bf G35} (2008)
  075001, [\href{http://arxiv.org/abs/0711.4022}{{\tt arXiv:0711.4022}}].

\bibitem{Bhattacharyya:2013rya}
G.~Bhattacharyya, D.~Das, P.~B. Pal, and M.~N. Rebelo, {\it {Scalar sector
  properties of two-Higgs-doublet models with a global U(1) symmetry}},  {\em
  JHEP} {\bf 10} (2013) 081, [\href{http://arxiv.org/abs/1308.4297}{{\tt
  arXiv:1308.4297}}].

\end{thebibliography}\endgroup
\end{document}